%% file: ms_nihao_budget.tex
\documentclass[useAMS,usenatbib]{mn2e}
\input psfig.sty
\usepackage{array}
\usepackage{amsmath}
\usepackage{amssymb}

\def \mean#1{\left< #1 \right>}

\def \apj {ApJ}
\def \apjl {ApJL}
\def \mnras {MNRAS}
\def \apjs {ApJS}
\def \aap {A\&A}

\def \na {NewA}

\def \spose#1{\hbox  to 0pt{#1\hss}}  
\def \lta{\mathrel{\spose{\lower 3pt\hbox{$\sim$}}\raise  2.0pt\hbox{$<$}}}
\def \gta{\mathrel{\spose{\lower  3pt\hbox{$\sim$}}\raise 2.0pt\hbox{$>$}}}
\def \ion#1#2{#1{\footnotesize{#2}}\relax}

\def \hi       {\ion{H}{I}}

\def \ovi      {\ion{O}{VI}}

\def \kms {\ifmmode  \,\rm km\,s^{-1} \else $\,\rm km\,s^{-1}  $ \fi }
\def \kpc {\ifmmode  {\rm kpc}  \else ${\rm  kpc}$ \fi  }  
\def \hkpc {\ifmmode  {h^{-1}\rm kpc}  \else ${h^{-1}\rm kpc}$ \fi  }  
\def \hMpc {\ifmmode  {h^{-1}\rm Mpc}  \else ${h^{-1}\rm Mpc}$ \fi  }  
\def \Msun {\ifmmode \rm M_{\odot} \else $\rm M_{\odot}$ \fi}
\def \hMsun {\ifmmode h^{-1}\,\rm M_{\odot} \else $h^{-1}\,\rm M_{\odot}$ \fi}
\def \hhMsun {\ifmmode h^{-2}\,\rm M_{\odot}\else $h^{-2}\,\rm M_{\odot}$ \fi}
\def \Lsun {\ifmmode L_{\odot} \else $L_{\odot}$ \fi} 
\def \hhLsun {\ifmmode h^{-2}\,\rm L_{\odot} \else $h^{-2}\,\rm L_{\odot}$ \fi}

\def\LCDM{$\Lambda$CDM }
\def \LCDM {\ifmmode \Lambda{\rm CDM} \else $\Lambda{\rm CDM}$ \fi}
\def \sig8 {\ifmmode \sigma_8 \else $\sigma_8$ \fi} 
\def \Omegam {\ifmmode \Omega_{\rm m} \else $\Omega_{\rm m}$ \fi} 
\def \Omegab {\ifmmode \Omega_{\rm b} \else $\Omega_{\rm b}$ \fi} 
\def \Omegar {\ifmmode \Omega_{\rm r} \else $\Omega_{\rm r}$ \fi} 
\def \fbar {\ifmmode f_{\rm b} \else $f_{\rm b}$ \fi} 
\def \OmegaL {\ifmmode \Omega_{\rm \Lambda} \else $\Omega_{\rm \Lambda}$\fi} 
\def \Deltavir {\ifmmode \Delta_{\rm vir} \else $\Delta_{\rm vir}$ \fi}
\def \rhocrit {\ifmmode \rho_{\rm crit} \else $\rho_{\rm crit}$ \fi}

\def \rs {\ifmmode r_{\rm s} \else $r_{\rm s}$ \fi} 
\def \rh {\ifmmode r_{\rm h} \else $r_{\rm h}$ \fi} 
\def \Rvir {\ifmmode R_{\rm vir} \else $R_{\rm vir}$ \fi}
\def \Vvir {\ifmmode V_{\rm  vir} \else  $V_{\rm vir}$  \fi} 
\def \Vmax {\ifmmode V_{\rm  max} \else  $V_{\rm max}$  \fi} 
\def \Mvir {\ifmmode M_{\rm  vir} \else $M_{\rm  vir}$ \fi}  
\def \Mhalo {\ifmmode M_{200} \else $M_{200}$ \fi}  
\def \Nvir {\ifmmode N_{\rm  vir} \else $N_{\rm  vir}$ \fi}  
\def \Jvir {\ifmmode J_{\rm vir} \else $J_{\rm vir}$ \fi} 
\def \Evir {\ifmmode E_{\rm vir} \else $E_{\rm vir}$ \fi} 
\def \lam {\ifmmode \lambda  \else $\lambda$ \fi} 
\def \lamp {\ifmmode \lambda^{\prime} \else $\lambda^{\prime}$  \fi} 
\def \lampc {\ifmmode \lambda^{\prime}_{\rm c} \else
  $\lambda^{\prime}_{\rm c}$  \fi} 

\def \xoff {\ifmmode x_{\rm off} \else $x_{\rm off}$ \fi}
\def \rhorms {\ifmmode \rho_{\rm rms} \else $\rho_{\rm rms}$ \fi}
\def \qbar {\ifmmode \bar{q} \else $\bar{q}$ \fi}

\def \Mb {\ifmmode M_{\rm b} \else $M_{\rm b}$ \fi} 
\def \eSF {\ifmmode \epsilon_{\rm SF} \else $\epsilon_{\rm SF}$ \fi} 
\def \Md {\ifmmode M_{\rm d} \else $M_{\rm d}$ \fi} 
\def \Mg {\ifmmode M_{\rm g} \else $M_{\rm g}$ \fi} 
\def \Rb {\ifmmode R_{\rm b} \else $R_{\rm b}$ \fi} 
\def \Rd {\ifmmode R_{\rm d} \else $R_{\rm d}$ \fi} 
\def \Rg {\ifmmode R_{\rm g} \else $R_{\rm g}$ \fi} 
\def \mgal {\ifmmode m_{\rm gal} \else $m_{\rm gal}$ \fi} 
\def \rj {\ifmmode {\cal R}_j \else ${\cal R}_j$ \fi} 
\def \lamgal {\ifmmode \lambda_{\rm gal} \else $\lambda_{\rm gal}$ \fi} 
\def \Vcirc {\ifmmode V_{\rm circ} \else $V_{\rm circ}$ \fi} 
\def \Vrot {\ifmmode V_{\rm rot} \else $V_{\rm rot}$ \fi} 
\def \Vflat {\ifmmode V_{\rm flat} \else $V_{\rm flat}$ \fi} 
\def \Mstar {\ifmmode M_{\rm star} \else $M_{\rm star}$ \fi} 
\def \Mgas {\ifmmode M_{\rm gas} \else $M_{\rm gas}$ \fi} 
\def \Mbar {\ifmmode M_{\rm bar} \else $M_{\rm bar}$ \fi}
\def \Rbar {\ifmmode R_{\rm bar} \else $R_{\rm bar}$ \fi} 

\def \DeltaIMF {\ifmmode \Delta_{\rm IMF} \else $\Delta_{\rm IMF}$ \fi}

\def \VV {\ifmmode V_{\rm 2.2}/V_{200} \else $V_{2.2}/V_{200}$ \fi} 
\def \dvr {\ifmmode \partial_{\rm VR} \else $\partial_{\rm VR}$ \fi} 

\title[Baryon Budget] {NIHAO VII: Predictions for the galactic baryon budget in dwarf to Milky Way mass haloes}

\author[Wang et al.]{Liang Wang$^{1,2,4}$\thanks{liang.wang@uwa.edu.au}, Aaron A. Dutton$^{3,4}$,
  Gregory S. Stinson$^4$, Andrea V. Macci\`o$^{3,4}$, 
\newauthor{Thales Gutcke$^4$, Xi Kang$^2$}\\
$^1$International Centre for Radio Astronomy Research (ICRAR)
, M468, University of Western Australia, 35 Stirling Hwy, \\
Crawley, WA 6009, Australia\\
$^2$Purple Mountain Observatory, the Partner Group of MPI f\"ur Astronomie, 2 West Beijing Road, Nanjing 210008, China\\
$^3$New York University Abu Dhabi, PO Box 129188, Abu Dhabi, UAE\\
$^4$Max-Planck-Institut f\"ur Astronomie, K\"onigstuhl 17, 69117 Heidelberg, Germany}
\begin{document}

\date{submitted to MNRAS}
             
\pagerange{\pageref{firstpage}--\pageref{lastpage}}\pubyear{2016}

\maketitle           

\label{firstpage}
             

\begin{abstract}
  We use the NIHAO galaxy formation simulations to make predictions
  for the baryonic budget  in present day galaxies ranging from dwarf
  ($\Mhalo\sim10^{10} \Msun$) to Milky Way ($\Mhalo\sim10^{12} \Msun$)
  masses.  The sample is made of 88 independent high resolution
  cosmological zoom-in simulations.  NIHAO galaxies reproduce key
  properties of observed galaxies, such as the stellar mass vs halo
  mass and cold gas vs stellar mass relations. Thus they make
  plausible predictions for the baryon budget.  We present the mass
  fractions of stars, cold gas ($T<10^4$K), cool gas ($10^4 < T <
  10^5$K), warm-hot gas ($10^5 < T < 5\times10^6$K), and hot gas (T$>
  5\times10^6$K), inside the virial radius, $R_{200}$.  Compared to
  the predicted baryon mass, using the dark halo mass and the
  universal baryon fraction, $f_{\rm b}\equiv \Omega_{\rm
    b}/\Omega_{\rm m}=0.15$, we find that all of our haloes are
  missing baryons. The missing mass has been relocated past 2 virial
  radii, and cool gas dominates the corona at low mass (M$_{200} \lta
  3 \times 10^{11} \Msun$) while the  warm-hot gas dominates at high
  mass (M$_{200} \gta 3 \times 10^{11} \Msun$).  Haloes of mass
  $\Mhalo\sim 10^{10}\Msun$ are missing $\sim 90\%$ of their baryons.
  More massive haloes ($\Mhalo\sim 10^{12}\Msun$) retain a higher
  fraction of their baryons, with $\sim 30\%$ missing, consistent with
  recent observational estimates.  Moreover, these more massive haloes
  reproduce the observed  fraction of cold, warm-hot and hot gas.  The
  fraction of cool gas we predict ($0.11\pm0.06$) is significantly
  lower than the observation from COS-HALOs (0.3-0.47), but agrees
  with the alternative analysis of \citet{Stern16}.
\end{abstract}


\setcounter{footnote}{1}


\section{Introduction}
\label{sec:intro}
Cosmic structure formation has redistributed the baryons from  a
nearly uniformly distributed plasma into a variety of states,
including stars, stellar remnants, cold (atomic and molecular) gas,
and hot (ionized) gas. The theories of galaxy formation can predict
the amount of mass in these different states, which can in turn be
tested by observational constraints.  

On cosmological scales, the ratio between the total baryonic and
gravitating mass is measured to be $f_{\rm b}\equiv \Omega_{\rm
  b}/\Omega_{\rm m}\simeq 0.15$ (The Planck Collaboration 2014).
However, the cold baryonic mass density implied by several galaxy
baryon estimates (mainly stars and cold gas) is only 3-8\% of
the big bang nucleosynthesis expectation \citep{Persic92, Fukugita98,
  Bell03, McGaugh10}.  The majority of the cosmic baryons are thought
to be in the form of hot gas around or between galaxies
\citep{Cen09}. Until recently only a fraction of these baryons had
been detected \citep{Bregman07, Shull12}.  This discrepancy is
referred to as the ``missing baryon problem''.  Several theoretical
studies with cosmological simulations have constrained the phase of
the potential reservoirs of the missing baryons in the
  intergalactic medium (IGM), and find a large fraction of the baryons
  with low density and high temperature resides between galaxies
\citep{Yoshida05, He05, Dave10, Zhu11, Haider16}.

As a part of the IGM, the circum galactic medium (CGM) is always 
treated as an major potential
reservoir of the missing baryons. \hi{} and metal absorption lines are
expected  to signpost such diffuse baryonic content. \hi{} is mainly
from gas with temperature T $\sim 10^4$ K so that it is able to detect
cold gas in the CGM.  Meanwhile, theoretical work has predicted that a
substantial  portion of the CGM is in the warm or hot phase with
temperature T $> 10^{4.5}$ K.  Gas enters this phase through
photoionization, accretion shocks or shocks caused by galactic winds
\citep{Voort12}. Such dilute halo gas is at T$\sim 10^{4.5-7}$ K, so
the detection is dominated by metal lines, e.g. \ovi{}.  Recent
advances in the detection of gas in the CGM have come  from the COS
survey \citep{Tumlinson11, Tumlinson13, Thom12,  Werk12, Werk13}.  On
the scale of Milky Way mass haloes $\Mhalo \sim 10^{12}\Msun$ a
significant amount of warm ($10^4 < T <5\times 10^6$K) gas has been detected
\citep{Werk14}, accounting for 33-88\% of the baryon budget. In the
future such observations will be extended to a wider range of halo
masses.  A number of large volume cosmological simulations
\citep{Ford13, Ford16, Suresh15, Oppenheimer16} and zoom-in
cosmological simulations \citep{Stinson12, Hummels13, Shull14} have
given  predictions for the \hi{} and \ovi{} absorption
lines. \citet{Gutcke16} compared the column density profile of \ovi{}
and \hi{} in the CGM of galaxies from the NIHAO \citep{Wang15}
cosmological hydrodynamical simulation suite with observations,
studied the covering fraction of dense \hi{}, looked at the shape of
the CGM and its chemical composition. The simulations
  reproduce  the observational covering fraction and column density
  profile of cool \hi{} well, and recover the observed trends of
  \ovi{} column density with luminosity and impact parameter. In
  common with other simulations, however, the column density of \ovi{}
  is lower and the extent of optically thick \hi{} is smaller than
  observed.

The physical properties of the CGM has been shown to be able to
test feedback models \citep{Sharma12,Marasco13}.  \citet{Dave09}
predicted galactic halo baryon fractions of galaxies with halo
masses ranging from $10^{11} \Msun$ to $10^{13} \Msun$ using
cosmological hydrodynamical simulations with a well-constrained
model for galactic outflows.  They found that, without the outflow
model, the baryon fraction inside the virial radius is roughly the
cosmic baryonic fraction, but with the outflow model, the baryon
fraction is increasingly suppressed in lower mass haloes.  By
comparing results at $z=3$ and $z=0$, they showed that large
haloes remove their baryons at early times while small haloes lose
baryons more recently due to the wind material taking longer to
return to low-mass galaxies than high-mass galaxies.
\citet{Muratov15} showed similar results that the gas and
baryon fractions are lower at lower redshift, after powerful
outflows at intermediate redshift $z \approx 0.5-2$ remove a large
amount of gas from the halo. Several simulations 
\citep[e.g.,][]{Crain07, Christensen16,Voort16} found the baryon fraction
and gas fraction are reduced compared to the cosmic baryon
fraction, especially in low mass haloes.
\citet{Sokolowska16} studied the halo gas of three Milky way-sized
galaxies using cosmological zoom-in simulations. They found that
most of missing baryons actually resides in warm-hot and hot gas
which contribute to 80\% of the total gas reservoir.  The recovered
baryon fraction within 3 virial radii is 90\%.  The warm-hot medium
is sensitive to the feedback model so that a reliable spatial
mapping of the warm-hot medium will provide  a stringent test for
feedback models.

In this paper we make predictions for the baryonic budget for stars,
cold, warm and hot gas in and around the virial radius of haloes of
mass ranging from $\Mhalo\sim 10^{10}\Msun$ to $10^{12}\Msun$. We use
a sample of 88 zoom-in galaxy formation simulations from the NIHAO
project.

Reproducing the stellar mass content in dark matter haloes
both today and in the past has been a formidable challenge for
cosmological galaxy formation simulations \citep{Weinmann12,
  Hopkins14}. Even the latest state-of-the art simulations have
trouble: the ILLUSTRIS simulation \citep{Vogelsberger14} strongly
overpredicts the stellar masses in dwarf galaxy haloes $(M_{200} \lta
10^{11}\Msun)$, while the EAGLE simulations \citep{Schaye15}
underpredict the peak of the star formation efficiency in halos of
mass $M_{200}\sim 10^{12}\Msun$. In contrast, the NIHAO galaxies are
consistent with the stellar mass vs halo mass relations from halo
abundance matching since redshift $z\sim 4$ \citep{Wang15}, the galaxy
star formation rate vs stellar mass relation since $z\sim 4$
\citep{Wang15}, and the cold gas mass vs stellar mass relation at
$z\sim 0$ \citep{Stinson15}.  Therefore, the simulations make
plausible predictions for the mass fractions and physical locations of
the warm and hot gas components.  We find that all the haloes contain
less baryons than expected according to the cosmic baryonic fraction,
but the missing fraction is strongly mass dependent.  

This paper is organized as follows: The cosmological hydrodynamical
simulations including star formation and feedback are briefly
described in  \S\ref{sec:sims}; In \S\ref{sec:budget} we present the
results including the baryonic budget, baryon distribution, and a
comparison with observations; \S\ref{sec:sum} gives a summary of our
results.

\section{Simulations} 
\label{sec:sims}

In this study we use simulations from the NIHAO (Numerical
Investigation of a Hundred Astrophysical Objects) project
\citep{Wang15}.   The initial conditions are created to keep the same
numerical resolution across the whole mass range with typically a
million dark matter particles inside the virial radius of the target
halo at  redshift $z=0$.  The halos to be re-simulated at higher
resolution with baryons have been extracted from 3 different pure
N-body simulations with a box size of 60, 20 and 15 $h^{-1}$ Mpc
respectively.  We adopted the  latest compilation of cosmological
parameters from the Planck  satellite \citep{Planck14}.   Dark
  matter particle masses range from $\sim 10^{4} \Msun$ in  our lowest
  mass haloes to $\sim 10^{6} \Msun$ in our most massive  haloes, and
  their force softenings range from $\sim$ 150 pc to $\sim$ 900 pc,
  respectively Gas particles are less massive by factor of
  $(\Omega_{\rm dm}/\Omega_{\rm b})\simeq 5.48$, and  the
  corresponding force softenings are 2.34 times smaller.  More
information on the collisionless parent simulations, the force
softenings and particle masses for the highest refinement level for
each simulation and sample  selection can be found in \citet{Dutton14}
and \citet{Wang15}.

We use the SPH hydrodynamics code {\sc gasoline} \citep{Wadsley04},
with a revised treatment of  hydrodynamics as described in
\citet{Keller14}.  The code includes a subgrid model for turbulent
mixing of metal and energy \citep{Wadsley08}, heating and cooling
include photoelectric heating of dust grains, ultraviolet (UV) heating
and ionization and  cooling due to hydrogen, helium and metals
\citep{Shen10}.  

The star formation and feedback modeling follows what was used in the
MaGICC simulations \citep{Stinson13}.   The gas is converted into
stars according to the Kennicutt-Schmidt Law when it satisfies a
temperature and density threshold. Stars feed both metals and energy
back into the ISM gas surrounding the region where they formed. SN
feedback is implemented using the blastwave formalism described in
\citet{Stinson06}. Pre-SN feedback is an attempt to consider radiation
energy from massive stars. Heating is introduced immediately after
massive stars form based on how much star light is radiated.  Our
simulations use thermal feedback to provide pressure support and
increase gas temperature above the star formation threshold, and thus
to decrease star formation.  There are two small changes in NIHAO
simulations compared to MaGICC: The change in  number of neighbors and
the new combination of softening length and  particle mass increases
the threshold for star formation from  9.3 to 10.3 cm$^{-3}$, the
increase of pre-SN feedback efficiency $\epsilon_{\rm ESF}$, from 0.1
to 0.13.  More details on the star formation and feedback modeling can
be found in \citet{Wang15}.

\begin{figure}
\centerline{
  \psfig{figure=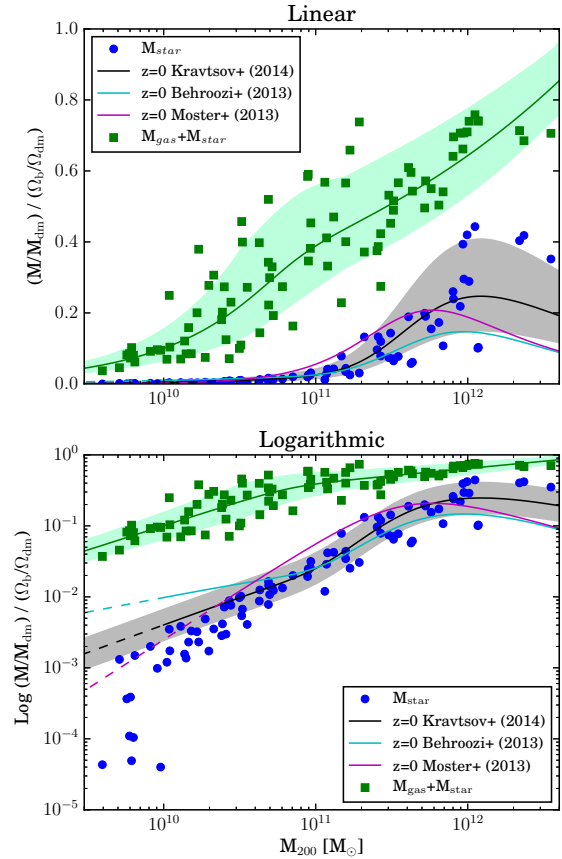,width=0.46\textwidth}
}
\caption{Fractional baryon content of our NIHAO simulations  as a
  function of halo mass. The green points show the ratio between the
  baryonic mass (stars + gas) inside the virial radius and the total
  baryonic mass associated with the dark matter halo. The blue points
  show the corresponding fraction for the stars. The solid green line
  and shaded region shows a double power-law fit, together with the
  1$\sigma$ scatter. For the stellar mass fraction we show several
  relations from halo abundance matching.    The linear (upper panel)
  and logarithmic (lower panel) scales emphasize the large amount of
  ``missing'' baryons, and the power-law nature of the relations,
  respectively.}
\label{fig:budget}
\end{figure}


\section{Baryon budget}
\label{sec:budget}

We define the fiducial baryonic mass as:
\begin{equation}
M_{\rm b} \equiv M_{\rm b}(R_{200})= \frac{f_{\rm b}}{1-f_{\rm b}}M_{\rm dm}(R_{200}) 
\label{equ:mb}
\end{equation}
where the $M_{\rm dm}$ is the total dark matter mass of the halo, and
the $f_{\rm b} = \Omegab/\Omegam \sim 0.15$ is the cosmic baryon
fraction (the ratio between baryon density and mass density including
baryonic mass plus dark matter), so that $M_{\rm b}$ would be the
baryonic mass inside the virial radius if the baryons followed the
dark matter closely.

Fig.~\ref{fig:budget} shows the ratio between the  mass of each baryon
component inside the virial radius  to the fiducial baryonic mass for
the most massive galaxy in each zoom-in region. We present the
fractions of total stellar mass within 20\% $R_{200}$ (blue
points),  and the total baryonic mass including stellar mass plus gas
mass (green points).  For the stellar mass fraction we also show the
relations from halo abundance  matching \citep{Moster13, Behroozi13,
  Kravtsov14}.  The grey area is the one sigma scatter around the mean
value  from \citet{Kravtsov14}.  Fig.~\ref{fig:budget} shows that all
haloes in our study contain less than the universal fraction of
baryons. The upper panel uses a linear y-axis scale, which highlights
the large amount of baryons that are missing, especially in low mass
haloes. The logarithmic scale in the lower panel highlights the
power-law nature of the relations.

The trends of each component fraction are similar, in that  the
fractions are relatively low in low mass haloes, and increase as the
halo mass increases.  The main difference between the different
components is the slope, with the baryonic mass fraction having a
shallower slope than the stellar mass fraction.  This is because in
low mass haloes ($\Mhalo\sim 10^{10}\Msun$) most of the baryons are in
the form of gas, while in the highest mass haloes we study
($\Mhalo\sim 10^{12}\Msun$) a substantial amount of gas
has been turned into stars.

Since most of the haloes we study are above the mass where the cosmic
UV background prevents gas from cooling, the missing baryons have most
likely been ejected from the central galaxies in supernova/stellar
feedback driven winds.  Although the lower mass galaxies have
converted a smaller fraction of their available baryons into stars,
and hence there is proportionally less energy available to drive an
outflow, they have expelled a larger fraction of their baryons.  This
is consistent with expectations from energy driven gas outflows, where
the lower star formation efficiency is more than compensated by the
shallower potential wells of lower mass halos according to the
mass loading factor and circular velocity  relation $\eta
\propto V^{-2}$ \citep[e.g.,][]{Dutton12, Christensen16}.

The behavior of the baryonic mass fraction, $f_{\rm bar}$, as a 
function of the halo mass is captured using a double power law formula:
\begin{equation}
\frac{f}{f_0} = \left( 
                        \frac{M_{200}}{\mathcal{M}_0} 
                        \right)^\alpha 
                        \left\{ 0.5 \left[ 1+\left( 
                        \frac{M_{200}}{\mathcal{M}_0} 
                        \right)^\gamma \right] 
                        \right\}^{\frac{\beta-\alpha}{\gamma}}.
\end{equation}
In this formula, the lower and higher mass ends have logarithmic slope
$\alpha$ and $\beta$, respectively, while $\gamma$ regulates how 
sharp the transition is from the lower to the higher ends.
Giving all points equal weight, the best fit parameters are as follows:
\begin{eqnarray}
\mathcal{M}_0  &=&  6.76 \times 10^{10} \nonumber \\
f_0  &=&   0.336 \nonumber \\
\alpha  &=&   0.684 \\
\beta  &=&   0.205 \nonumber \\
\gamma  &=&   3.40\nonumber  
\end{eqnarray}

  The three most massive galaxies in the NIHAO suite have fairly high
  stellar masses compared to abundance matching results (see blue
  filled circles in Fig.~\ref{fig:budget}), and are thus possibly
  overcooled.  However, they don't significantly bias the baryon
  fraction fitting formula since the high mass slope ($M_{200} >
  10^{11}\Msun$) is constrained by $\sim 40$ other haloes.

The green shaded region indicates the scatter about the best fit line,
which is 0.151 dex for haloes with mass in the range of  
$3\times 10^{9}\Msun < M_{200} <2\times 10^{10} \Msun$, 0.236 for
halo mass in 
$2\times 10^{10}\Msun < M_{200} < 7\times 10^{10} \Msun$,
0.125 for halo mass in
$7\times 10^{10}\Msun < M_{200} < 3\times 10^{11} \Msun$
and 0.052 for halo mass in
$3\times 10^{11}\Msun < M_{200} < 3.5\times 10^{12} \Msun$.

As might be expected, haloes with the highest masses  we study ($\sim
10^{12}\Msun$) have high baryon fractions ($\sim 0.7$) with relatively
small scatter (0.05 dex). As halo mass decreases, the baryon fraction
decreases and the scatter increases reaching a maximum of $\sim 0.24$
dex in haloes of mass $\sim 4\times10^{10}\Msun$.  However, at the
lowest halo masses we study, below $10^{10}\Msun$, the scatter starts
to decrease. By contrast the scatter in the stellar mass fraction
increases below this scale. Thus it seems unlikely that stellar
feedback is primarily responsible for the low baryon
fractions. Rather, we suggest an increased importance of the UV
background, which heats gas to above the virial temperature, thus
preventing it from collapsing into the low mass haloes. The
baryon fractions in low mass haloes are thus controlled primarily 
by the halo masses, and thus are independent of the large scatter in the
stellar mass.

\begin{figure*}
\centerline{
  \psfig{figure=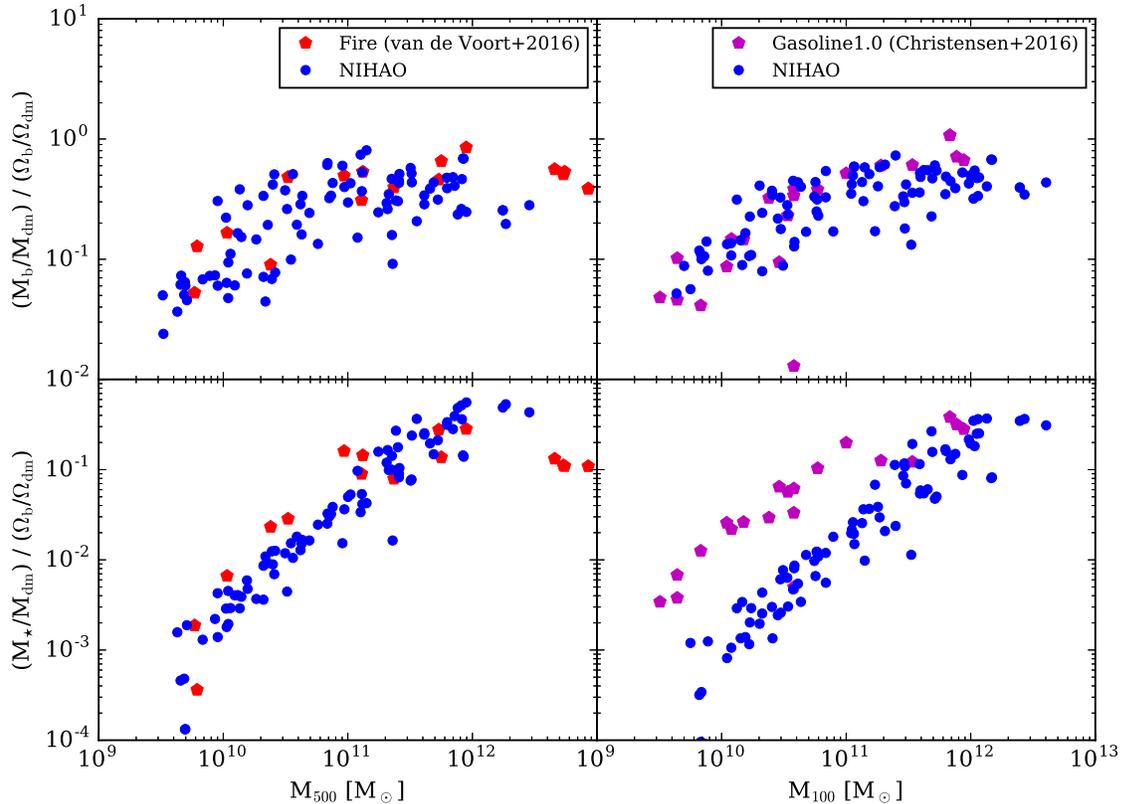,width=0.85\textwidth}
}
\caption{Comparison between baryon (upper panels) and stellar
    fractions (lower panels) in the NIHAO simulations (blue circles)
    with simulations from \citet{Voort16} (left, red pentagons) and
    \citet{Christensen16} (right, purple pentagons).}
\label{fig:fire_com}
\end{figure*}

\subsubsection{Comparison with other zoom-in simulations}
  In Fig.~\ref{fig:fire_com} we compare the NIHAO results for the
  stellar and baryonic mass fractions with other recent
  state-of-the-art zoom-in simulations from \citet{Voort16} and
  \citet{Christensen16}.  Since these authors both use different halo
  mass definitions than we adopt here, we re-calculate the NIHAO
  results using a virial radius defined at $500\times$ critical
  density (left panels) and $100\times$ critical density (right
  panels).  The simulations analyzed in \citet{Voort16} are from the
  FIRE project \citep{Hopkins14}, which uses a different
  hydrodynamical code and different sub-grid model for star formation
  and feedback. The simulations analyzed in \citet{Christensen16} use
  a similar code as NIHAO, i.e.,  {\sc gasoline}, but with important
  differences. NIHAO uses an upgraded version that improves mixing
  \citep{Keller14}, and includes stronger feedback which was found to
  be necessary to delay star formation at early times
  \citep{Stinson13}. Despite these different codes employed, the
  baryon fractions in the three sets of simulations are in remarkably
  good agreement.

For the massive haloes with  M$_{500} > 10^{12} \Msun$, there are
three galaxies from NIHAO and four from FIRE.  There is a hint that
more massive haloes have lower baryon fractions.  Of course the caveat
here is that neither NIHAO or FIRE includes AGN feedback, which could
have a significant impact on the gaseous content of these haloes.

For the stellar fractions, \citet{Voort16} shows almost the same
result as our work. Even at the low mass end, their two lowest mass
galaxies may suggest the  large scatter as found in NIHAO.  However,
\citet{Christensen16} predicts one magnitude higher stellar fractions
than what we find for halo  masses below $10^{11.5}
\Msun$. This difference is likely due to the additional early stellar
feedback included in the NIHAO version of gasoline. 

In summary, while the stellar mass fractions are dependent on the
sub-grid models, the baryon fractions (for haloes in the mass range
$10^{10}\lta M_{200} \lta 10^{12}\Msun$) appear insensitive to the
details of the simulation code. In particular all three codes predict a
greater fraction of missing baryons in lower mass haloes. 
It suggests that the baryon fraction correlates 
strongly with halo mass, due to the deeper potential well of the halo and/or stronger 
ram pressure experienced by the outflowing gas. However, Fig.~\ref{fig:fire_com}
 suggests that the 
 specific implementation of the feedback model does not matter, as long as the feedback 
is efficient in driving outflows.

\begin{figure*}
\centerline{
  \psfig{figure=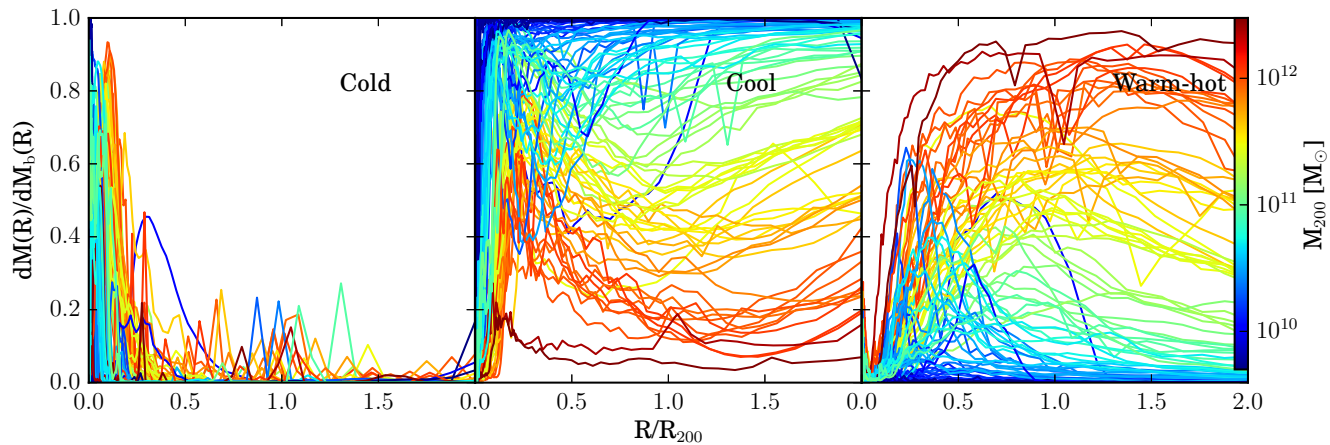,width=1.0\textwidth}
}
\caption{Radial profile of the mass fraction of the gas in each phase
         to total baryonic mass in each radial bin 
         at $z=0$ for all galaxies in NIHAO sample.
         Each solid line is from one galaxy and colour coded with
         the halo mass.}
\label{fig:corona}
\end{figure*}

\subsection{Mass budget of the corona}
\label{sec:corona}

In Fig.~\ref{fig:corona}, we present the radial distribution of
gas in different phases at $z=0$, normalized to the total baryonic
mass profile, such that, in the region far away from the central galaxy
where the stars are rare, for a given halo the four phases add up to
unity, each line is colour coded by the virial mass.  
All simulations share a common attribute.  The cold gas (T$<10^4$K)
is mostly located  near the center  ($R<0.2R_{200}$)
where most stars in galaxies form.  In contrast, the cool ($10^4$K$<$T$<10^5$K)
and  warm-hot ($10^5$K$<$T$<5\times10^6$K) gas are
located at large distances with roughly constant fractions up to 2
times $R_{200}$.  The hot gas (T$>5\times10^6$K) is a minority component
for all galaxies in the NIHAO sample, with the maximum hot gas
fraction at any radius being  less than 5\%.

Despite these similarities, we find a considerably higher proportion
of cool gas in lower mass galaxies (M$_{200} < 10^{11} \Msun$) in the
whole corona region.  For higher mass galaxies, warm-hot gas dominates
the corona  which signals stronger virial shocks and higher
efficiency  of feedback.  Even beyond the virial radius, the cool and
warm-hot gas has similar features as the gas within virial radius
which reveals the gas surrounding galaxies within large distances is
the major reservoir of baryons.

\subsection{Where are the missing baryons?}
\label{sec:where}

\begin{figure}
\centerline{
  \psfig{figure=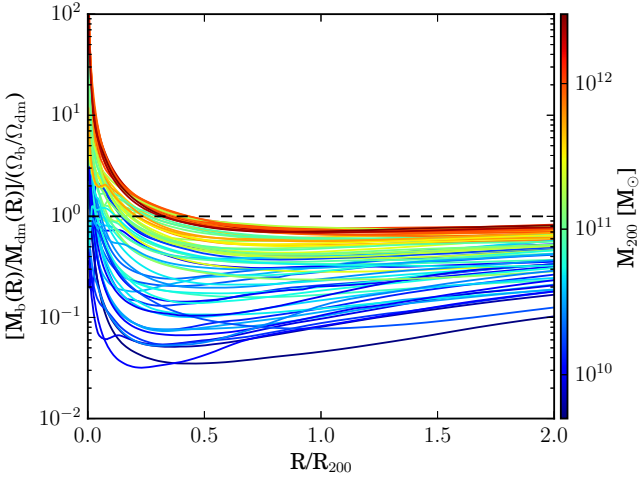,width=0.5\textwidth}
}
\caption{Baryon distribution of each galaxy from NIHAO simulations. 
 The lines are colour coded by their halo mass, which shows a
  clear trend that the more massive haloes preserve more baryons at
  all radii.}
\label{fig:rps}
\end{figure}

Fig.~\ref{fig:rps} shows the cumulative fraction of total baryons  for
each simulation. Here the y-axis is the ratio between the baryonic to
dark matter mass, $M_{\rm b}(<R) / M_{\rm dm}(<R)$, enclosed
within a sphere of radius, $R$, normalized by the cosmic
baryon-to-dark matter ratio, $\Omega_{\rm b}/\Omega_{\rm dm}$.

Each solid curve represents a halo, and the curves are coloured by
their halo mass (red for high masses to blue for low masses).  Broadly
speaking, the curves have a similar shape, with a normalization that
depends on halo mass. They have a cusp in  the central region where
the stars and cold gas dominate, then become flat in the outer region.
More massive haloes have higher baryon fractions at all radii.  At
small radii, the baryon to dark matter ratio is higher than the cosmic
value due to gas dissipation. Beyond 0.5 virial radii, all haloes
  are missing baryons.  Even beyond the virial radius, there is
little change in the baryon fraction up to 2 virial radii.  We thus
conclude that the missing baryons are well outside of the virial
radius.

 \citet{Ford16} compare results for the cumulative baryon profile
  from two cosmological hydrodynamic simulations which employed
  different prescriptions for galactic  outflow models. They
  constructed samples of simulated galaxies with a similar
  distribution of stellar masses to that of COS-Halos.

  In the hybrid energy/momentum driven winds model (``ezw'') 65\% of
  all available baryons are inside the halo. This is broadly
  consistent with estimates of baryonic mass derived from
  \citet{Werk14} and our  finding in Fig.~\ref{fig:budget}. Comparing
  the cumulative baryon fraction profile in detail, Fig.10 in
  \citet{Ford16} shows the baryon fraction within 0.1$\Rvir$ is only
  $\sim 30\%$ and the fraction  gradually increases to 65\% at
  $\Rvir$. In the NIHAO simulations the profile of galaxies with halo
  masses below $10^{11} \Msun$ in Fig.~\ref{fig:rps} have  similar
  features, the more massive galaxies all have roughly flat slopes.

The simplified constant wind outflow model (``cw'') shows
a lower fraction at all radii inside $\Rvir$, even though the ``ezw'' 
and ``cw'' models generally gives similar observational absorption 
line properties. This suggests that the cumulative baryon
fraction profile is complementary to the total amount of CGM gas for
distinguishing between competing outflow models.

\begin{figure}
\centerline{
  \psfig{figure=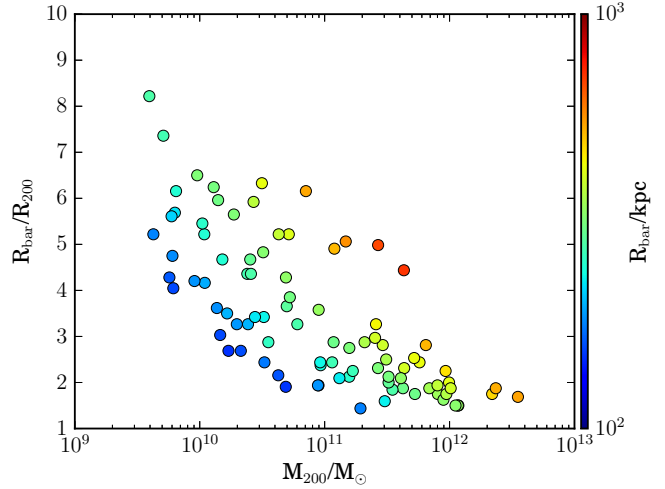,width=0.5\textwidth}
}
\caption{Normalized baryon radius as function of total virial mass.
         The points are colour coded by  
         baryon radius of each galaxy.}
\label{fig:missvr}
\end{figure}

To estimate how far the baryons escape, we measured the radius,
$\Rbar$, within which the total baryon mass equals the fiducial
baryonic mass defined by Eq.~\ref{equ:mb}. This is a lower limit to
the true extent of the missing baryons since the baryon mass includes
gas and stars that belong to nearby lower mass haloes.  It is a
  radius within the high-resolution volume of the simulations since
  the mass fraction of low resolution dark matter particles
  in a shell between 0.9 and  1.0 $\Rbar$
  shows most NIHAO galaxies have a fraction close to 0, with largest
  only around 10 - 20\%.  Fig.~\ref{fig:missvr} shows the baryon
radius of each galaxy as function of the virial mass. In physical
units, we find that the baryon radius generally increases with virial
mass. When normalized by the virial radius, the distance baryons are
ejected gradually decreases as halo mass increases, varying from
$\Rbar/R_{200}\sim 5$ at a halo mass of $M_{200}\sim 10^{10}\Msun$ to
$\sim 2$ at a halo mass of $M_{200}\sim 10^{12}\Msun$.

\begin{figure*}
\centerline{
  \psfig{figure=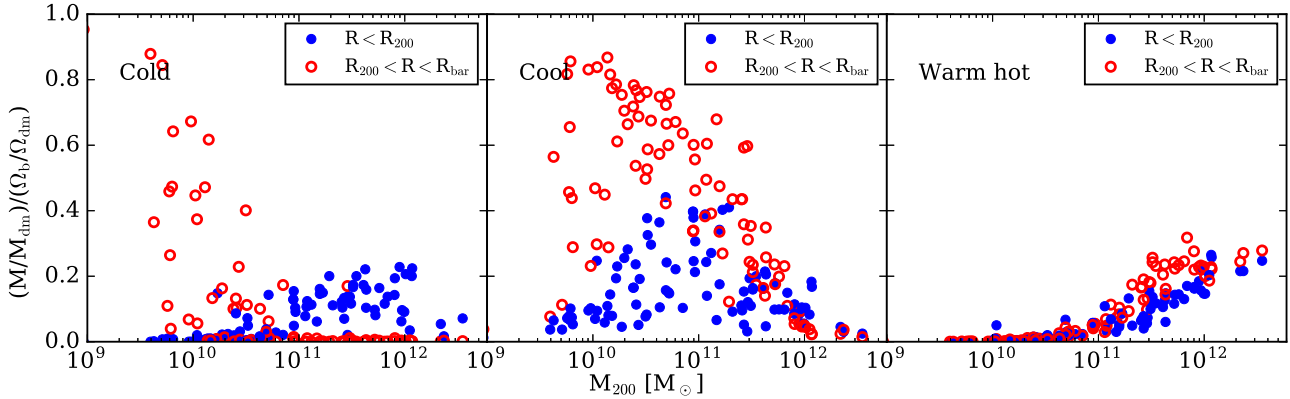,width=1.0\textwidth}
}
\caption{ Mass fraction of gas in three phases (relative to the
    fiducial baryonic mass within the virial radius) inside the
    virial radius (filled blue points) and between  virial and baryon
    radius (open red points), respectively. Cool gas is the dominant
    component of the fiducial baryonic mass for most galaxies
    ($10^{10} < M_{200}/\Msun < 10^{12}$). Cold  and warm-hot gas is
    the majority only for galaxies at the low and high mass
    ends, respectively.}
\label{fig:inout}
\end{figure*}

In Fig.~\ref{fig:inout}, we show the mass fractions of gas (in three
phases) inside the virial radius (filled blue points) and between the
virial and baryon radius (open red points).  All fractions are
  relative to the fiducial baryonic mass within the  virial radius.
The fractions of cold gas, $f_{\rm cold}$, are shown in the left
panel.  Inside the virial radius $f_{\rm cold}$ increases gradually
from zero at a halo mass of $10^{10}\Msun$ to $\sim 20\%$ at a halo
mass of $10^{12}\Msun$.  The gas outside the virial radius has the
opposite and much stronger trend: $f_{\rm cold}\sim 50\%$ in haloes of
mass $10^{10}\Msun$ and decreases to zero by halo masses of
$10^{11}\Msun$.
The fractions of cool gas, $f_{\rm cool}$, are shown in the middle
panel.  Inside the virial radius, $f_{\rm cool}$, has a maximum of
40\% at a halo mass of $10^{11} \Msun$, and declines to less than 10\%
below and above halo masses of $10^{10}\Msun$ and $10^{12}\Msun$,
respectively.  For most haloes there is more cool gas outside than
inside the virial radius.  In haloes of mass $10^{10}\Msun$, $f_{\rm
  cool}\sim 80\%$, and decreases to less than 10\% by a halo mass of
$10^{12}\Msun$.
The fractions for warm-hot gas,$f_{\rm warm}$,  are shown in the
right panel.  The trends of the gas inside and outside the virial
radius are quite similar, $f_{\rm warm}$ increases monotonically with
halo mass with maximum values of $\sim 30\%$.
The hot gas isn't shown since it is negligible both
inside and outside the virial radius across the whole mass range we
study.
We thus conclude that, for galaxies with halo masses in the range
$10^{10} \lta M_{200}\lta 10^{11} \Msun$, the majority of baryons
associated with the dark matter halo are in the cold and cool phases,
and are located well outside of virial radius.  For haloes in the mass
range $10^{11} \lta M_{200}\lta 10^{12} \Msun$, the fractions of cold
gas, cool gas and  warm-hot gas are comparable.

\subsection{Comparison with observations
of Milky Way mass haloes}

Observations can gain information of CGM from absorption and emission
lines. Although emission lines allow us to directly obtain a 3D
picture of the distribution of gas in the CGM, emission line studies
preferentially probe the dense gas closer to galaxies, since gas
emissivity scales with the square of density. While the situation is
improving with new facilities, e.g.~\citet{Hayes16}, absorption lines
are  the most common observational constraints on the physical state
of the CGM.

\begin{figure*}
\centerline{
  \psfig{figure=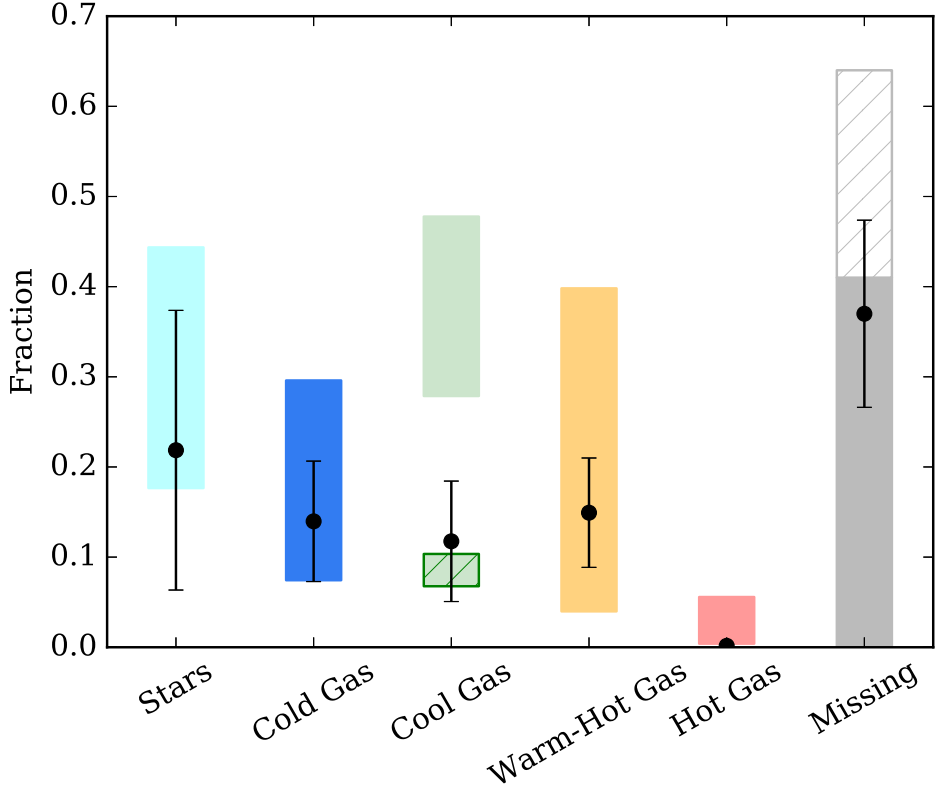,width=0.8\textwidth}
}
\caption{Baryonic budget  of NIHAO haloes of mass $3.5\times 10^{11} <
  M_{200}/\Msun < 3.5 \times 10^{12}$ (black points  with 1$\sigma$
  error bars) compared with observations of  $M_{200} \sim
  10^{12}\Msun$ haloes (shaded regions). There is good agreement,
  except for the cool gas which has two conflicting measurements:
  \citet[][upper]{Werk14} upper, \citet[][lower hatched]{Stern16}.  }
\label{fig:comparison}
\end{figure*}

The COS-HALOs survey is filling in details about the CGM at low
redshift \citep{Peeples14, Tumlinson11, Tumlinson13, Werk12, Werk13,
  Werk14}.  For the CGM of low-redshift $\sim L^*$ galaxies
($\Mstar\sim 10^{10.5}\Msun$), \citet{Tumlinson13} and
\citet{Peeples14} constrain the mass of  the warm-hot CGM ($T \sim
10^{5-6.7}$K), \citet{Werk14} provides a strict lower limit to the
mass of cool material ($T \sim 10^{4-5}$K) in the CGM of these
galaxies.  In a study using X-rays, \citet{Anderson13} place a
constraints on the mass of  hot gas ($T > 5\times10^6$K) residing in
the extended hot halos.  These observational constraints are shown in
Fig.~\ref{fig:comparison} with the same colour scheme as in Fig.11 of
\citet{Werk14}.  The stellar mass fractions are based on halo
abundance matching as described in \citet{Kravtsov14}, while the cold
disk gas mass comes from \citet{Dutton11}.  The upper limits to
  the missing fraction are calculated using the lower limits to all
  the fractions. There are two upper limits shown which correspond to
  the two measurements of the cool-gas fraction
  \citep{Werk14,Stern16}. The lower limit to the missing fraction is
  consistent with zero.

The black points and error bars show the mean values and standard
deviation of the mass fraction of stars and different components of
gas in our most massive galaxies ($3.49 \times 10^{11} \Msun$ $<$
M$_{200}$ $<$  $3.53 \times 10^{12} \Msun$).  The gas is assigned to a
range of temperature bins:  cold gas (T $< 10^4$ K), cool gas ($10^4$
K $<$ T $< 10^5$ K),  warm gas ($10^5$ K $<$ T $<5\times 10^6$ K) and hot gas
(T $>5\times 10^6$ K).  The observations and the simulations match well in
every phase except the cool CGM gas, where the observations from
\citet{Werk14} find $3\times$ the mass that NIHAO simulations predict.

If the observations are correct, the simulations have either ejected
cool gas too far, or  they have created a CGM with the wrong mix of
gas temperatures. The total gas fractions (0.39 in COS-HALOs, 0.41 in
NIHAO) suggest the latter option.   However, \citet{Stern16} developed
a new method to constrain the physical conditions in the cool CGM from
measurements of ionic  columns densities. This new method combines the
information available from different sight-lines during the
photoionization modeling, and was applied to the COS-HALOs data
yielding a total cool CGM mass within the virial radius of
$1.3\times10^{10}\Msun$.  The corresponding cool gas fraction is shown
by the green hashed bar in Fig~\ref{fig:budget} and is in good
agreement with our prediction.  We should note that \citet{Stern16}
does not follow a same  assumption on CGM ionization that assume all
\ovi{} is produced through photoionization and no collisional
ionization, however,  \citet{Gutcke16} finds the CGM in $L^*$ galaxies
in NIHAO is mostly  ionized by collisional ionization.

 Since \citet{Crain07} and \citet{Voort13} show that  soft X-ray
  emission traces gas with temperatures a few times the  virial
  temperature (T$_{\rm vir} = 10^6$ K for M$_{200} \sim  10^{12}
  \Msun$), we also estimate the hot gas fraction with lower boundaries
  ($3\times 10^6$ K and $10^6$ K) between the hot gas and  warm-hot
  gas. The hot gas fractions increase to 0.00511 ($3\times 10^6$K)
  and 0.0417 ($10^6$ K), while the warm-hot gas fractions decrease by
  a corresponding amount. Even with these different definitions the
  NIHAO simulations are in good agreement with the observed warm-hot
  and hot gas fractions.

As the Fig.~\ref{fig:corona} and Fig.~\ref{fig:inout} show, the cool
gas is the most important component for most haloes in the NIHAO
  simulations ($10^{10} < M_{200}/\Msun < 10^{12}$). The cold and
  warm-hot gas only dominate for galaxies at the lower and higher mass
  ends.  As the CGM of lower mass galaxies will soon be observed,
  Table~\ref{tab:comparison} lists information about CGM mass
  fractions of the different components of gas in haloes down to a
  halo mass of $\sim 10^{10}\Msun$.

  A recent estimation for the baryonic fractions from a set of eight
  Milky Way-sized zoom-in cosmological simulations \citep{Colin16} is
  generally consistent with our results. The fraction of cool gas
  from their simulations is even lower  ($f_{\rm cool}$ = 0.034) than
  ours and the result from  \citet{Stern16}. On the other hand,
  \citet{Peeples14} gave  a conservative observational estimate about
  warm-hot gas and showed that the fraction of gas in this phase is
  only 5\%,  which is the lower limit of the estimates from
  \citet{Werk14}  and less than the average values from
  \citet{Colin16}  ($f_{\rm warm-hot} \approx 24\%$) and our finding
  ($f_{\rm warm-hot} \approx 17\%$). Therefore, the mix of
  temperatures of the CGM has large uncertainties. More accurate
knowledge of the physical properties of the CGM are necessary to
better constrain feedback models and to understand the role of the CGM
in galaxy formation.

\section{Summary}
\label{sec:sum}

We have used the NIHAO galaxy simulation suite \citep{Wang15} to study
the statistical features of the baryonic budget and distribution
spanning halo masses of $\sim 10^{10}$ to $\sim 10^{12}\Msun$. NIHAO
is a large (currently 88) set of high resolution cosmological zoom-in
hydrodynamical galaxy formation simulations. As shown in previous
papers the NIHAO galaxies reproduce several key observed scaling
relations, and thus they make plausible predictions for the
  baryon budget in and around galaxies.  We summarize our results as
follows:

\begin{itemize}
\item All of the NIHAO haloes have a lower baryon to dark matter
  ratio,  inside the virial radius, than the cosmic baryon fraction
  (Fig.~\ref{fig:budget}). We refer to the cosmic baryon fraction associated 
  with each dark matter halo as the fiducial baryons.

\item Lower mass haloes are missing a larger fraction of their fiducial baryons,
  even though they convert a much lower fraction of the baryons into
  stars (Fig.~\ref{fig:budget}).  Similar trends are found by
    other recent simulations \citep{Christensen16,Voort16} using
    different codes (Fig.~\ref{fig:fire_com}).
 
\item The missing baryons have been expelled well beyond the
  virial radius, $R_{200}$, (Fig.~\ref{fig:rps}). Relative to the
  virial radius, the baryons are expelled to smaller radii in more
  massive haloes: $\Rbar \sim 5 R_{200}$ for $M_{200}=10^{10}\Msun$ and
  $\Rbar \sim 2 R_{200}$ for $M_{200}=10^{12}\Msun$
  (Fig.~\ref{fig:missvr}).

\item Cold gas ($T<10^4$K) is mostly restricted to be within 0.2
  virial radii (Fig.~\ref{fig:corona}). Cool gas ($10^4 < T < 10^5$K)
  dominates the baryonic mass outside the virial radius, as well
    as outside 20\% $R_{200}$, at low masses ($M_{200}\lta 3\times
  10^{11} \Msun$) while the warm-hot gas ($10^5 < T <5\times 10^6$K)
  dominates at high masses ($M_{200}\gta 3\times 10^{11} \Msun$)
  (Figs.~\ref{fig:corona} \& \ref{fig:inout}). 

\item For the highest mass haloes in our study $\sim 10^{12}\Msun$ our
  simulations are consistent with the observed  fractions
  \citep[e.g.][]{Werk14} of stars, cold gas, warm and hot gas inside
  the virial radius (Fig.~\ref{fig:comparison}). 

\item For the cool gas we predict $f_{\rm cool}=0.11\pm0.06$
      which is significantly lower than the observations from COS-HALOs
      ($f_{\rm cool}=0.28-0.48$), but is in excellent agreement with the
      analysis of \citet{Stern16}.
      
\end{itemize}

 We hope our results will motivate observers to obtain more
  accurate measurements of the mass fractions in different phases of
  the CGM over a wide range of galaxy masses, and simulators to make
  the corresponding predictions.

\begin{table*}
  \caption{The baryonic budget parameters for NIHAO galaxies
    in different halo  mass bins. We refer to gas in the temperature range  T
  $<$ $10^4$ K as cold; $10^4$ K $\leqslant$ T $<$ $10^5$ K as cool;
  $10^5$ K $\leqslant$ T $<$ $5\times10^6$ K as warm;  and T $\geqslant$
  $5\times10^6$ K as hot.}
\begin{center}
\begin{tabular}{ccccc}
\hline
\input{budgettable4.txt}
\hline
\end{tabular}
\label{tab:comparison}
\end{center}
\end{table*}

\section*{Acknowledgments} 

We thank the two anonymous referees whose suggestions greatly  improve
the paper.  We thank Freeke van de Voort for kindly sharing data
  of the FIRE simulation.  {\sc Gasoline} was written by Tom Quinn
and James Wadsley. Without their contribution, this paper would have
been impossible.
The simulations were performed on the {\sc theo} cluster of the
Max-Planck-Institut f\"ur Astronomie and the {\sc hydra} cluster at
the Rechenzentrum in Garching; and the Milky Way supercomputer, funded
by the Deutsche Forschungsgemeinschaft (DFG) through Collaborative
Research Center (SFB 881) "The Milky Way System" (subproject Z2),
hosted and co-funded by the J\"ulich Supercomputing Center (JSC). We
greatly appreciate the contributions of all these computing
allocations.
AAD, GSS and AVM acknowledge support through the
Sonderforschungsbereich SFB 881 “The Milky Way System” (subproject A1)
of the German Research Foundation (DFG).  The analysis made use of the
pynbody package \citep{Pontzen13}.
The authors acknowledge support from the MPG-CAS through the
partnership programme between the MPIA group lead by AVM and the PMO
group lead by XK.
LW acknowledges support of the MPG-CAS student programme.
XK acknowledge the support from 973 program (No. 2015CB857003,
2013CB834900), NSFC project No.11333008 and the ``Strategic Priority
Research Program the Emergence of Cosmological Structures'' of the
CAS(No.XD09010000).



\appendix

\section{Resolution test}

\begin{figure*}
\centerline{
  \psfig{figure=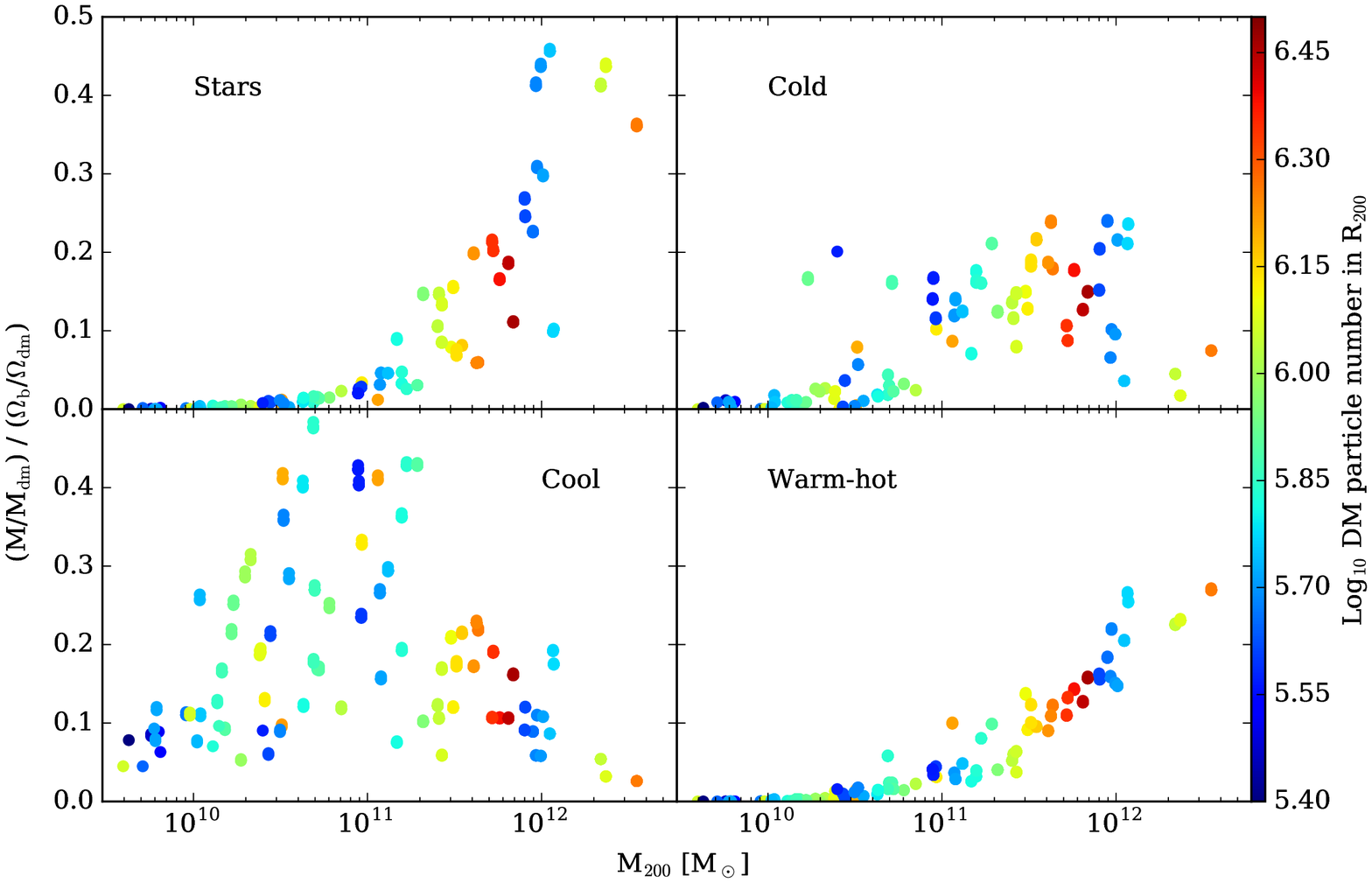,width=0.95\textwidth}
}
\caption{Baryon content of NIHAO simulations as a function of
    halo mass color coded by number of dark matter particles per
    halo. This shows the stars and gas content of the simulations is
    insensitive to order of magnitude changes in particle
    resolution.}
\label{fig:budget_res}
\end{figure*}

 Fig.~\ref{fig:budget_res} shows the budget fraction of stars, cold, cool
  and warm-hot gas color coded by the number of dark matter particles
  inside the virial radius. Since one can imagine that with increased
  resolution, higher densities can be achieved in the CGM, potential
  leading to more cooling and larger cold/cool gas fractions.  In our
  simulations we find no dependence of stars and gas fractions with particle
  number, indicating these quantities are not sensitive to numerical
  resolution using the NIHAO sub-grid model.

\label{lastpage}
\end{document}

%% file: budgettable4.txt
$\mean{\log_{10}(M_{200}/\Msun)}$ & $9.974\pm 0.211$  & $10.563\pm0.148$ & $11.182\pm0.229$ & $11.879\pm0.218$\\
\hline
$\mean{\log_{10}(M_{\rm b}/\Msun)}$ &$9.225\pm0.208$ &$9.805\pm0.140$ &$10.409\pm0.229$ &$11.093\pm0.211$\\
$\mean{M_{\star}/M_{\rm b}}$ &$1.64\times10^{-3}$ &$8.71\times10^{-3}$ &$5.39\times10^{-2}$ & 0.219 \\ 
$\sigma_{\star}$ &$1.63\times10^{-3}$  & $5.06\times10^{-3}$  & $4.43\times10^{-2}$  & 0.155  \\ 
$\mean{M_{\rm cold}/M_{\rm b}}$ &$1.32\times10^{-2}$ &$3.88\times10^{-2}$ & 0.118 & 0.140 \\ 
$\sigma_{\rm cold}$ &$5.77\times10^{-3}$  & $4.08\times10^{-2}$  & $3.60\times10^{-2}$  & $6.68\times10^{-2}$  \\ 
$\mean{M_{\rm cool}/M_{\rm b}}$ & $9.53\times 10^{-2}$ & 0.195 & 0.233 & 0.118 \\ 
$\sigma_{\rm cool}$ &$5.47\times10^{-2}$  & 0.135 &0.147 &$6.68\times10^{-2}$  \\ 
$\mean{M_{\rm warm}/M_{\rm b}}$ &$1.03\times10^{-3}$ &$1.40\times10^{-2}$ &$4.87\times10^{-2}$ & 0.146 \\ 
$\sigma_{\rm warm}$ &$1.28\times10^{-3}$  & $7.36\times10^{-3}$  & $3.04\times10^{-2}$  & $5.22\times10^{-2}$  \\ 
$\mean{M_{\rm hot}/M_{\rm b}}$ &0.000 &0.000 &0.000 &$5.11\times10^{-3}$  \\ 
$\sigma_{\rm hot}$ &0.000 &0.000 &0.000 &$4.53\times10^{-3}$  \\ 
$\mean{M_{\rm missing}/M_{\rm b}}$ & 0.889 & 0.743 & 0.545 & 0.370 \\ 
$\sigma_{\rm missing}$ &$3.66\times10^{-2}$  & 0.137 &0.123 &0.104  \\ 

%% file: ms_nihao_budget.bbl
\begin{thebibliography}{}


\bibitem[Anderson et al.(2013)]{Anderson13} Anderson, M.~E., Bregman, J.~N., Dai, X.\ 2013, \apj, 762, 106

\bibitem[Agertz et al.(2007)]{Agertz07} Agertz, O., Moore, B., Stadel, J.\ 2007, \mnras, 380, 963


\bibitem[Behroozi et al.(2013)]{Behroozi13} Behroozi, P.~S.,
  Wechsler, R.~H., \& Conroy, C.\ 2013, \apj, 770, 57

\bibitem[Bell et al.(2003)]{Bell03} Bell, E.~F., McIntosh, D.~H., Katz, N., Weinberg, M.~D.,\ 2003, \apj, 585, 117

\bibitem[Bregman (2007)]{Bregman07}
Bregman, J.~N.\ 2007, ARAA, 45, 221


\bibitem[Cen \& Ostriker (1999)]{Cen09} 
Cen, R.~Y., Ostriker, J.~P.\ 1999, \apj, 514, 1

\bibitem[Christensen et al.(2016)]{Christensen16}
Christensen, C.~R., Dav{\'e}, R., Governato, F., et al.\ 2016, \apj, 824, 57

\bibitem[Colin et al.(2016)]{Colin16}
Colin, P., Avila-Reese, V., Roca-Fabrega, S., et al.\ 2016, \apj, in press

\bibitem[Crain et al.(2007)]{Crain07}
Crain, R.~A., Eke, C.~R., Frenk, C.~S., et al.\ 2007, \mnras, 377, 41


\bibitem[Dav{\'e} (2009)]{Dave09} Dav{\'e}, R.\ 2009, ASPC, 419, 347D

\bibitem[Dav{\'e} et al.(2010)]{Dave10} Dav{\'e}, R., Oppenheimer, B.~D., Katz, N., et al.\ 2010, \mnras, 408, 2051

\bibitem[Dutton et al.(2011)]{Dutton11} Dutton, A.~A., Conroy, 
  C., van den Bosch, F.~C., et al.\ 2011, \mnras, 416, 322
  
\bibitem[Dutton(2012)]{Dutton12} Dutton, A.~A.\ 2012,
  \mnras, 424, 3123
  
\bibitem[Dutton \& Macci{\`o}(2014)]{Dutton14} Dutton,
  A.~A., \& Macci{\`o}, A.~V.\ 2014, \mnras, 441, 3359 



    

%
\bibitem[Ford et al.(2013)]{Ford13} Ford, A.~B., Oppenheimer, B.~D., Dav{\`e}, R., et al.\ 2013, \mnras, 432, 89

%
\bibitem[Ford et al.(2016)]{Ford16} Ford, A.~B., Werk, J.~W., Dav{\`e}, R., et al.\ 2016, \mnras, 459, 1745

\bibitem[Fukugita et al.(1998)]{Fukugita98} Fukugita, M., Hogan, C.~J., Peebles, P.~J.~F.\ 1998, \apj, 503, 518


\bibitem[Gutcke et al.(2016)]{Gutcke16} Gutcke, T.~A., Stinson, G.~S., Macci{\`o}, A.~V., et al.\ 2016, arXiv:1602.06956, MNRAS in press



\bibitem[Haider et al.(2016)]{Haider16}
Haider, M., Steinhauser, D., Vogelsberger, M., et al.\ 2016, \mnras, 457, 3024

\bibitem[Hayes et al.(2016)]{Hayes16}
Hayes, M., Melinder, J., {\"O}stlin, G., et al.\ 2016, arXiv:1606.04536

\bibitem[He et al.(2005)]{He05}
He, P., Feng, L.~L., Fang, L.~Z.\ 2005, \apj, 623, 601

\bibitem[Hopkins et al.(2014)]{Hopkins14} Hopkins, P.~F., Kere{\v s}, D., O{\~n}orbe, J., et al.\ 2014, \mnras, 445, 581 

%
\bibitem[Hummels et al.(2013)]{Hummels13} Hummels, C.~B., Bryan, G.~L., Smith, B.~D., et al.\ 2013, \mnras, 430, 1548




\bibitem[Keller et al.(2014)]{Keller14} Keller, B.~W., Wadsley, 
  J., Benincasa, S.~M., \& Couchman, H.~M.~P.\ 2014, \mnras, 442, 3013

\bibitem[Kravtsov et al.(2014)]{Kravtsov14} Kravtsov, A., 
Vikhlinin, A., \& Meshscheryakov, A.\ 2014, arXiv:1401.7329 

  



\bibitem[Marasco et al.(2013)]{Marasco13} 
Marasco, A., Marinacci, F., Fraternali, F.\ 2013, \mnras, 433, 1634

\bibitem[McGaugh et al.(2010)]{McGaugh10} McGaugh, S.~S., 
Schombert, J.~M., de Blok, W.~J.~G., Zagursky, M.~J.\ 2010, \mnras,
708, 14

\bibitem[Moster et al.(2013)]{Moster13} Moster, B.~P., Naab, T., 
\& White, S.~D.~M.\ 2013, \mnras, 428, 3121 

\bibitem[Muratov et al.(2015)]{Muratov15} Muratov, A.~L., Ker{\v e}s, D., Faucher-Gigu{\`e}re, C., et al.\ 2015, \mnras, 454, 2691




\bibitem[Oppenheimer et al.(2016)]{Oppenheimer16} Oppenheimer, B.~D., Crain, R.~A., Schaye, J., et al.\ 2016, \mnras, 460, 2157 


\bibitem[the Planck Collaboration et  al.(2014)]{Planck14}
  Planck Collaboration, Ade, P.~A.~R., Aghanim, N., et al.\ 2014,
  \aap, 571, AA16 
 
\bibitem[Peeples et al.(2014)]{Peeples14} Peeples, M.~S., Werk, J.~K., Tumlinson, J., et al.\ 2014, \apj, 786, 54

\bibitem[Persic \& Salucci(1992)]{Persic92} Persic, M., Salucci, P.\ 1992, \mnras, 258, 14

\bibitem[Pontzen et al.(2013)]{Pontzen13} Pontzen, A., Ro{\v s}kar, R., Stinson, G., \& Woods, R.\ 2013, Astrophysics Source Code Library, 1305.002 





\bibitem[Schaye et al.(2015)]{Schaye15} Schaye, J., Crain,
  R.~A., Bower, R.~G., et al.\ 2015, \mnras, 446, 521

  
\bibitem[Shen et al.(2010)]{Shen10} Shen, S., Wadsley, J., 
\& Stinson, G.\ 2010, \mnras, 407, 1581 

\bibitem[Sharma et al.(2012)]{Sharma12}
Sharma, P., McCourt, M., Parrish, I.~J., Quataert, E.\ 2012, \mnras, 427, 1219

\bibitem[Shull et al.(2012)]{Shull12}
Shull, J.~M., Smith, B.~D., Danforth, C.~W.\ 2012, \apj, 759, 23

%
\bibitem[Shull (2014)]{Shull14} Shull, J.~M.\ 2014, \apj, 784, 142

\bibitem[Sokolowska et al.(2016)]{Sokolowska16}
Sokolowska, A., Mayer, L., Babul, A., Madau, P., Shen, S.\ 2016, \apj, 819, 21

\bibitem[Stern et al.(2016)]{Stern16} Stern, J., Hennawi, J.~F., Prochaska, J.~X., \& Werk, J.~K.\ 2016, \apj, 830, 87 


\bibitem[Stinson et al.(2006)]{Stinson06} Stinson, G.~S., Seth, A., Katz, N., et al.\ 2006, \mnras, 373, 1074

%
\bibitem[Stinson et al.(2012)]{Stinson12} Stinson, G.~S., Brook, C., Prochaska, J.~X., et al.\ 2012, \mnras, 425, 129

\bibitem[Stinson et al.(2013)]{Stinson13} Stinson, G.~S., Brook, 
C., Macci{\`o}, A.~V., et al.\ 2013, \mnras, 428, 129 

\bibitem[Stinson et al.(2015)]{Stinson15} Stinson, G.~S., Dutton, A.~A., Wang, L., et al.\ 2015, \mnras, 454, 1105 

\bibitem[Suresh et al.(2015)]{Suresh15} Suresh, J., Rubin, K.~H.~R., Kannan, R., et al.\ 2015, arXiv:1511.00687


\bibitem[Thom et al.(2012)]{Thom12}
Thom, C., Tumlinson, J., Werk, J.~K.\ 2012, \apjl, 758, L41

\bibitem[Tumlinson et al.(2011)]{Tumlinson11} Tumlinson, J., Thom, C., Werk, J., et al.\ 2011, Science, 334, 948

\bibitem[Tumlinson et al.(2013)]{Tumlinson13} Tumlinson, J., Thom, C., Werk, J., et al.\ 2013, \apj, 777, 59


\bibitem[van de Voort \& Schaye (2012)]{Voort12} van de Voort, F., Schaye, J.\ 2012, \mnras, 423, 2991

\bibitem[van de Voort \& Schaye (2013)]{Voort13} van de Voort, F., Schaye, J.\ 2013, \mnras, 430, 2688

\bibitem[van de Voort et al.(2016)]{Voort16} van de Voort, F., Quataert, E., Hopkins, P.~F., et al.\ 2016, \mnras, 463, 4533 

  
\bibitem[Vogelsberger et al.(2014)]{Vogelsberger14} Vogelsberger, M., 
Genel, S., Springel, V., et al.\ 2014, \mnras, 444, 1518 



\bibitem[Wadsley et al.(2004)]{Wadsley04} Wadsley, J.~W., Stadel, 
J., \& Quinn, T.\ 2004, \na, 9, 137 

\bibitem[Wadsley et al.(2008)]{Wadsley08} Wadsley, J.~W., 
Veeravalli, G., \& Couchman, H.~M.~P.\ 2008, \mnras, 387, 427 

\bibitem[Wang et al.(2015)]{Wang15} Wang, L., Dutton, A.~A.,  Stinson, G.~S., et al.\ 2015, \mnras, 454, 83

  \bibitem[Weinmann et al.(2012)]{Weinmann12} Weinmann, S.~M., Pasquali, A., Oppenheimer, B.~D., et al.\ 2012, \mnras, 426, 2797 

\bibitem[Werk et al.(2012)]{Werk12} Werk, J.~k., Prochaska, J.~X., Thom, C., et al.\ 2012, \apjs, 198, 3

\bibitem[Werk et al.(2013)]{Werk13} Werk, J.~k., Prochaska, J.~X., Thom, C., et al.\ 2013, \apjs, 204, 17

\bibitem[Werk et al.(2014)]{Werk14} Werk, J.~k., Prochaska, J.~X., Thom, C., et al.\ 2014, \apj, 792, 8



\bibitem[Yoshida et al.(2005)]{Yoshida05} 
Yoshida, N., Furlanetto, S.~R., Hernquist, L.\ 2005, \apj, 618L, 91


\bibitem[Zhu et al.(2011)]{Zhu11} Zhu, W., Feng, L.~L., Fang, L.~Z.\ 2011, \mnras, 415, 1093

\end{thebibliography}
